\documentclass[a4paper,10pt]{article}
\usepackage{graphicx}
\usepackage{comment}
\usepackage{amssymb}
\usepackage{amsmath}
\usepackage{nicefrac}
\usepackage{graphicx}
\usepackage{dcolumn}
\usepackage{bm}
\usepackage{comment}
\usepackage{multirow}
\usepackage{cite}
\usepackage{xcolor}
\usepackage{url}
\usepackage{mathrsfs}

\begin{document}

\newcommand{\bin}[2]{\left(\begin{array}{c}\!#1\!\\\!#2\!\end{array}\right)}
\newcommand{\threej}[6]{\left(\begin{array}{ccc}#1 & #2 & #3 \\ #4 & #5 & #6 \end{array}\right)}
\newcommand{\sixj}[6]{\left\{\begin{array}{ccc}#1 & #2 & #3 \\ #4 & #5 & #6 \end{array}\right\}}
\newcommand{\regge}[9]{\left[\begin{array}{ccc}#1 & #2 & #3 \\ #4 & #5 & #6 \\ #7 & #8 & #9 \end{array}\right]}
\newcommand{\La}[6]{\left[\begin{array}{ccc}#1 & #2 & #3 \\ #4 & #5 & #6 \end{array}\right]}
\newcommand{\hj}{\hat{J}}
\newcommand{\hux}{\hat{J}_{1x}}
\newcommand{\hdx}{\hat{J}_{2x}}
\newcommand{\huy}{\hat{J}_{1y}}
\newcommand{\hdy}{\hat{J}_{2y}}
\newcommand{\huz}{\hat{J}_{1z}}
\newcommand{\hdz}{\hat{J}_{2z}}
\newcommand{\hup}{\hat{J}_1^+}
\newcommand{\hum}{\hat{J}_1^-}
\newcommand{\hdp}{\hat{J}_2^+}
\newcommand{\hdm}{\hat{J}_2^-}

\huge

\begin{center}
A new approach to include electron interaction effects in super transition array opacity theory
\end{center}

\vspace{0.5cm}

\large

\begin{center}
Daniel Aberg$^{a}$, Brian G Wilson$^{a,}$\footnote{wilson9@llnl.gov} and
Jean-Christophe Pain$^{b,c,}$\footnote{jean-christophe.pain@cea.fr}
\end{center}

\normalsize

\begin{center}
\it $^a$Lawrence Livermore National Laboratory, P.O. Box 808, L-414, Livermore, California 94551, USA\\
\it $^b$CEA, DAM, DIF, F-91297 Arpajon, France\\
\it $^c$Universit\'e Paris-Saclay, CEA, Laboratoire Mati\`ere en Conditions Extr\^emes,\\
\it 91680 Bruy\`eres-le-Ch\^atel, France\\

\end{center}

\vspace{0.5cm}

\begin{abstract}
A method is presented for the improved calculation of super-shell partition functions which include the repulsive electron-electron interaction energy terms in the Boltzmann factor. Heretofore these interaction terms were approximately treated via the use of Feynman-Jensen inequalities. Such investigations are of particular interest for the super-transition-array approach of hot-plasma radiative opacity.
\end{abstract}

\section{Introduction} 

The super-transition-array (STA) statistical method \cite{BARSHALOM1989} has been shown to be rather powerful for the modeling of hot dense plasmas in local thermodynamic equilibrium (LTE). It enables one to take into account the huge number of populated electron configurations. Central to the method is the computation of ionization distributions by calculating the partition function
\begin{equation}
{U_\Xi } = \sum\limits_{\vec p \in \Xi } {G\left( {\vec p} \right){e^{ - \beta \,\left\{ {E\left( {\vec p} \right) - \mu \cdot Q_\Xi} \right\}}}}
\end{equation}
over a restricted ensemble of configurations specified by the total number of electrons in groupings of atomic orbitals (super-shells),
denoted as the super-configuration $\Xi$. Here $\beta = {1 \mathord{\left/ {\vphantom {1 {kT}}} \right. \kern-\nulldelimiterspace} {kT}}$ is the inverse temperature, $\mu$ is the chemical potential, and ${Q_\Xi }$ is the sum of the super-shell occupations. The statistical weight for each configuration $\vec p$ (denoted as a vector of sub-shell occupations) within the super-configuration is given by the product of binomial coefficients
\begin{equation}
G\left( {\vec p} \right) = \prod\limits_{i = 1}^N {\left( {\begin{array}{*{20}{c}}
{{g_i}}\\
{{p_i}}
\end{array}} \right)},
\end{equation}
where ${g_i}$ is the integer degeneracy of sub-shell ``$i$'' ($0 \le {p_i} \le {g_i}$). For a central-field Hamiltonian the configuration-average total energy is given by
\begin{equation}\label{econf}
E\left( {\vec p} \right) = \sum\limits_{i = 1}^N {{p_i}{\varepsilon _i}} + \frac{1}{2}\sum\limits_{i,j = 1}^N {{p_i}\left( {{p_j} - {\delta _{ij}}} \right){\Delta _{ij}}},
\end{equation}
where the ${\varepsilon _i}$ (energy of sub-shell $i$) and ${\Delta _{ij}}$ (interaction energy between sub-shells $i$ and $j$) are evaluated from the central-field Hamiltonian, and ${\delta _{ij}}$ is Kronecker's delta symbol. 

In the STA formalism, a super-transition-array, corresponding to a set of transition arrays associated with a given single-electron jump, is represented by a Gaussian distribution, whose first moments (intensity, average energy and variance) are expressed, as explained below, using partition functions in the canonical thermodynamic ensemble (the number of electrons in a super-configuration being fixed).

The average value of any quantity $A\left( {\vec p} \right)$ over all the configurations of $\Xi$ reads
\begin{equation}\label{avera}
\left\langle A \right\rangle = \frac{1}{{{U_\Xi }}}\;\sum\limits_{\vec p \in \Xi } {G\left( {\vec p} \right){e^{ - \beta \,\left\{ {E\left( {\vec p} \right) - \mu Q_\Xi} \right\}}}A\left( {\vec p} \right)}.
\end{equation}
All relevant quantities needed for opacity calculations, such as average occupations, transition array strengths, centroids and widths, can be calculated from them. \cite{BARSHALOM1995}

The sums contained in the above formulas run usually over a very large number of electron configurations.
A direct evaluation of these expressions is therefore a hard and sometimes an almost impossible task.
However, when the quadratic dependence with respect to the populations of orbitals (due to two-body interactions)
is neglected then the partition function of the super-configuration factorizes into a product of partition functions for each super-shell
\begin{equation}\label{facto}
{U_\Xi } = \prod\limits_{\Phi \in \Xi } {U\left[ {{Q_\Phi },\vec g} \right]},
\end{equation}
where ${{Q_\Phi }}$ is the integer occupation of super-shell $\Phi$ and the super-shell partition functions are of the form
\begin{equation}\label{parto}
U\left[ {{Q_\Phi },\vec g} \right] \equiv \mathop {\sum\limits_{{p_1} = 0}^{{g_1}} {\sum\limits_{{p_2} = 0}^{{g_2}} {...} } }\limits_{{p_1} + {p_2} + ...{p_N} = {Q_\Phi }} \sum\limits_{{p_N} = 0}^{{g_N}} {\left( {\begin{array}{*{20}{c}}
{{g_1}}\\
{{p_1}}
\end{array}} \right)\left( {\begin{array}{*{20}{c}}
{{g_2}}\\
{{p_2}}
\end{array}} \right)...\left( {\begin{array}{*{20}{c}}
{{g_N}}\\
{{p_N}}
\end{array}} \right)X_1^{{p_1}}X_2^{{p_2}}...X_N^{{p_N}}},
\end{equation}
with
\begin{equation}
{X_i} = {e^{ - \beta \left( {{\varepsilon _i} - \mu } \right)}}.
\end{equation}
In our compact notation it is to be implicitly understood that only the sub-set of ${X_i}$ and ${\varepsilon _i}$ corresponding to the $N$ sub-shells in the super-configuration need be included.

Taking into account electron-electron interactions, i.e. the quadratic term in the expression of the energy of a configuration (see Eq. (\ref{econf})), does not allow us to factorize the partition function sub-shell by sub-shell, as in Eqs. (\ref{facto}) and (\ref{parto}). In the original STA method by Bar-Shalom, Oreg and Klapisch \cite{BARSHALOM1989}, this term is replaced, in the Boltzmann factors, by its average over the configurations belonging to the super-configuration under consideration. This results in a statistics of the ``independent-electron'' type, with a correction resulting from simplified treatment of the interaction terms.

This independent electron approximation allows recursion relations \cite{GILLERON2004,WILSON2007} and other techniques \cite{PAIN2023,WILSON2022,PAIN2020} for their rapid and exact evaluation. In large part this is possible because using
the identity \cite{GILLERON2004,WILSON1999}
\begin{equation}
{\delta _{0,L}} = \frac{1}{{2\pi i}}\int\limits_{ - i\pi + \lambda }^{ + i\pi + \lambda } {dt\;{e^{tL}}} \quad \mathrm{with}\quad L = Q - \sum\limits_{i = 1}^N {{p_i}},
\end{equation}
allows one to construct a closed form generating function \cite{FAUSSURIER1999,PAIN2011}
\begin{equation}
{U_Q}\left[ {\vec g} \right] = \frac{1}{{2\pi i}}\oint {dz\frac{{\Im \left( z \right)}}{{{z^{Q + 1}}}}} \quad \mathrm{where}\quad \Im \left( z \right) = \prod\limits_{i = 1}^N {{{\left( {1 + z{X_i}} \right)}^{{g_i}}}}.
\end{equation}

Different attempts were made to approximate the interaction terms beyond the original way \cite{BARSHALOM1989}. For instance, saddle-point approximations were widely used (see the end of section \ref{sec3}). Faussurier, Wilson and Chen \cite{FAUSSURIER2001} proposed a decomposition into a reference electron system and a first-order correction. The procedure proves to be highly efficient for evaluating both the free energy and orbital populations. Later on, it was suggested to extend that approach by incorporating a reference energy accounting for the interaction between two electrons within a given sub-shell \cite{PAIN2009}. This extension was made possible through an efficient recursion relation for calculating partition functions \cite{GILLERON2004}.

\section{Approximating interacting electrons}\label{sec3} 

In order to re-introduce the effects of two-body interactions, one proceeds by introducing a ``reference'' independent electron energy expression \cite{FAUSSURIER2001}
\begin{equation}
{E^0}\left( {\vec p} \right) = \sum\limits_{i = 1}^N {{p_i}\,\varepsilon _i^0}
\end{equation}
and what is termed a ``first order correction energy''
\begin{equation}
{E^1}\left( {\vec p} \right) \equiv E\left( {\vec p} \right) - {E^0}\left( {\vec p} \right).
\end{equation}
One thus has
\begin{equation}
{U_\Xi } = {\left\langle {{e^{ - \beta \,{E^1}}}} \right\rangle _0}\;U_\Xi ^0,
\end{equation}
where for any $A\left( {\vec p} \right)$ 
\begin{equation}
{\left\langle A \right\rangle _0} \equiv \frac{1}{{U_\Xi ^0}}\;\sum\limits_{\vec p \in \Xi } {G\left( {\vec p} \right)\,{e^{ - \beta \left\{ {{E^0}\left( {\vec p} \right) - \mu \cdot {Q_\Xi }} \right\}}}\,A\left( {\vec p} \right)}
\end{equation}
and
\begin{equation}
\left\langle A \right\rangle = \frac{{{{\left\langle {{e^{ - \beta {E^1}}}A} \right\rangle }_0}}}{{{{\left\langle {{e^{ - \beta {E^1}}}} \right\rangle }_0}}}.
\end{equation}

The basic assumptions of super-transition array theory are

(1) The super-shell populations, being proportional to ${U_Q}\left[ {\vec g} \right]$ are easily approximated by invoking Jensen's inequality \cite{PAIN2009,GILLERON2011}
\begin{equation}
{\left\langle {{e^{ - \beta {E^1}}}} \right\rangle _0} \ge {e^{ - \beta {{\left\langle {{E^1}} \right\rangle }_0}}}.
\end{equation}
An important benefit of using an independent electron reference system is that the evaluation of the right hand side is factorized in super-shell partition functions, as the expression for ${E^1}$ is at most quadratic in occupation variables, and averages over a product of sub-shells belonging to two different super-shells itself factorizes.

Note that we have flexibility in the choice of the reference single particle energies in order to optimally mitigate the effect of the Feynman-Jensen approximation, and this optimization can be super-configuration dependent.

(2) Many observables (e.g. array widths, etc.) can employ the approximation
\begin{equation}
{\left\langle {{e^{ - \beta {E^1}}}A} \right\rangle _0} \simeq {\left\langle {{e^{ - \beta {E^1}}}} \right\rangle _0}{\left\langle A \right\rangle _0}\quad \Rightarrow \quad \left\langle A \right\rangle  \simeq {\left\langle A \right\rangle _0}
\end{equation}
thus limiting the need to evaluate higher order correlations in population averages.

\section{Additional treatments of interdependent electrons} 

Early attempts \cite{GREEN1964} at incorporating inter-electron repulsion employed an an operator technique which provides a simple representation of the exact generating function. If one makes use of the algebra of these operators, one can evolve simple and exact
symbolic expressions for $\left\langle {{p_i}} \right\rangle$ and for $\left\langle {{p_i}{p_j}} \right\rangle - \left\langle {{p_i}} \right\rangle \left\langle {{p_j}} \right\rangle$. Indeed, it becomes apparent that one can, in principle, find symbolic expressions for any higher moment of the $p's$. The symbolic expressions, however, depend upon complicated averages which, themselves, are almost
impossible to compute exactly.

Another previously attempted approach aims to convert the interacting electron Boltzmann factor into an independent electron system via the introduction of auxiliary fields by using the Hubbard-Stratonovich transformation \cite{WILSON1993,HAZAK2012,KURZWEIL2013,KURZWEIL2016}
\begin{equation}
{e^{\frac{1}{2}\alpha {p^2}}} = \int\limits_{ - \infty }^{ + \infty } {dx\;{e^{ - \pi {x^2}}}{e^{\sqrt {2\pi \alpha } \,x\,p}}}.
\end{equation}
Here, an additional auxiliary field ${x_j}$ is included for each diagonal interaction term ${\Delta _{ii}}p_i^2$ and two auxiliary fields for each off diagonal interaction term (noting ${\Delta _{ij}} = {\Delta _{ji}}$) in the form
\begin{equation}
{\Delta _{ij}}{p_i}{p_j} = \frac{1}{4}{\Delta _{ij}}\left\{ {{{\left( {{p_i} + {p_j}} \right)}^2} - {{\left( {{p_i} - {p_j}} \right)}^2}} \right\}.
\end{equation}
This results in the closed form expression of ${N^2}$ auxiliary fields:
\begin{equation}
\int\limits_{ - \infty }^{ + \infty } {...} \int\limits_{ - \infty }^{ + \infty } {d\left\{ {{x_{ij}}} \right\}\;{e^{ - \pi \sum\limits_{ij} {x_{ij}^2} }}\prod\limits_k {{e^{ - \beta \left\{ {{{\tilde \varepsilon }_k} - \mu } \right\}{p_k}}}} },
\end{equation}
where
\begin{equation}
{\tilde \varepsilon _k} = {\varepsilon _k} - \frac{1}{2}{x_{kk}} + \sum\limits_m {{\lambda _{km}}{x_{km}}}
\end{equation}
with ${\lambda _{ij}} = i\sqrt {2\pi \beta {\Delta _{ij}}}$. As before, the product over independent electron terms (now a complex valued function of the auxiliary fields) can be performed under occupation constraints.

The problem with this approach is, apart from the high dimensionality of the auxiliary fields, that each auxiliary degree of freedom is a continuous variable that must be sampled over an infinite domain. Thus to get accurate results, many evaluations must be
performed, and so the utility of this approach has been limited to approximate saddle point evaluations of the multidimensional integral\cite{FAUSSURIER1997,FAUSSURIER1999}.

\section{Discrete Hubbard-Stratonovich Transformations} 

The infinite domain of the multi-dimensional integral produced from the traditional Hubbard-Stratonovich transformation ultimately arises because such a formula is valid for all real values of occupation $p$, whereas in our intended application only integer values
of $0 \le p \le g$ (where $g$ is even) are required. Discrete Hubbard-Stratonovich transformations of the form
\begin{equation}
{e^{\frac{1}{2}\alpha {p^2}}} = \sum\limits_{m = 1}^{\left( {{g \mathord{\left/
 {\vphantom {g 2}} \right.
 \kern-\nulldelimiterspace} 2}} \right) + 1} {{\omega _m}{e^{ - {x_m}p}}} = \sum\limits_{m = 1}^{\left( {{g \mathord{\left/
 {\vphantom {g 2}} \right.
 \kern-\nulldelimiterspace} 2}} \right) + 1} {{\omega _m}{{\left( {{\phi _m}} \right)}^p}}
\end{equation}
with specific sets of weights ${\omega _m}$ and roots ${\phi _m}$ (and thus ${x_m}$) can be made exact for $p = 0,1,2,...,g$ (as well as the unphysical value $p=g+1$) by solving the set of $2g$ nonlinear equations in $2g$ unknowns (roots and weights).

Closed form solutions for the roots and weights for systems with orbital degeneracy were first presented in Ref. \cite{GUNNARSSON1997} for $g=2$
and $g=4$. These authors also presented (for the cases $g=2,4$) alternative forms by pre-specifying one of the unknowns
and abandoning the exactness for $p=g+1$.

Analytic non-linear solution sets are easily obtained through $g=6$ with computer algebraic programs such as {\sc Mathematica} \cite{MATHEMATICA}, but higher orbital degeneracies could not be obtained with {\sc Mathematica} in closed form, nor easily solved for numerically.

 In order to obtain solutions for $g=8$ and higher we turn to the formal theory of Gaussian Quadratures and orthogonal polynomials \cite{NUMERICALRECIPES}, where
\begin{equation}
\int {W\left( \phi \right)} f\left( \phi \right)d\phi \simeq \sum\limits_{k = 1}^N {{\omega _k}f\left( {{\phi _k}} \right)}
\end{equation}
is exact if $f\left( \phi \right)$ is a polynomial.  The roots ${\phi _k}$ are obtained as the zeros of the polynomial
${P_N}\left( \phi \right)$, which are obtained by the recursive relations
\begin{equation}
{P_{j + 1}}\left( \phi \right) = \left( {\phi  - {a_j}} \right){P_j}\left( \phi \right) - {b_j}{P_{j - 1}}\left( \phi \right)\quad \mathrm{with} \quad {P_{ - 1}}\left( \phi \right) = 0\quad \mathrm{and}\quad {P_0}\left( \phi \right) = 1,\quad j = 0,1,2,...
\end{equation}
with
\begin{equation}
{a_j} = \frac{{\left\langle {x{P_j}} \right|\left. {{P_j}} \right\rangle }}{{{c_j}}},\quad \quad {b_j} = \frac{{{c_j}}}{{{c_{j - 1}}}}\quad \mathrm{and}\quad {c_j} = \left\langle {{P_j}} \right|\left. {{P_j}} \right\rangle
\end{equation}
where
\begin{equation}
\left\langle f \right|\left. g \right\rangle \equiv \int {W\left( \phi \right)f\left( \phi \right)g\left( \phi \right)},
\end{equation}
while the weights are given by
\begin{equation}\label{26}
{\omega _j} = \frac{{{c_{N - 1}}}}{{{P_{N - 1}}\left( {{\phi _j}} \right){{P'}_N}\left( {{\phi _j}} \right)}}.
\end{equation}
We now define $\Gamma \equiv {e^{{\alpha \mathord{\left/ {\vphantom {\alpha 2}} \right. \kern-\nulldelimiterspace} 2}}}$, which is less than (/greater than) unity for repulsive (/attractive) interactions. 
With a suitable transformation of real variables for attractive interactions ($\Gamma \ge 1$), we have from the original continuous Hubbard-Stratonovich integral all the moments
\begin{equation}
\int {W\left( \phi \right){\phi ^p}} = {\Gamma ^{{p^2}}},
\end{equation}
which together with known analytic solutions for the roots for small values of $N$ allows us to work backwards to deduce the
recursion coefficients
\begin{equation}
{a_j} = {\Gamma ^{4j + 1}} + {\Gamma ^{4j - 1}} - {\Gamma ^{2j - 1}}\quad \mathrm{and} \quad b_j = {\Gamma ^{6j - 4}}\left\{ {{\Gamma ^{2j}} - 1} \right\},\quad \quad j = 0,1,2,...
\end{equation}
Invoking analytic continuation assures the use of these same polynomials for repulsive ($\Gamma < 1$) interactions. It now becomes apparent why no analytic solutions exist for $g=8$ and higher: analytic solutions are possible for ${P_N}=0$ when quadratic, cubic, or quartic, but in general not for quintics or higher degree.

\section{Obtaining Roots and Weights} 

Rather than work directly with the Gaussian Quadrature polynomials
\begin{equation}
\begin{array}{l}
{p_0}\left( x \right) = 1\\
{p_1}\left( x \right) = x - \Gamma \\
{p_2}\left( x \right) = {x^2} - {\Gamma ^3}\left( {1 + {\Gamma ^2}} \right)x + {\Gamma ^6}\\
{p_3}\left( x \right) = {x^3} - {\Gamma ^5}\left( {1 - \Gamma + {\Gamma ^2}} \right)\left( {1 + \Gamma + {\Gamma ^2}} \right){x^2} + {\Gamma ^{10}}\left( {1 - \Gamma + {\Gamma ^2}} \right)\left( {1 + \Gamma + {\Gamma ^2}} \right){x^3} - {\Gamma ^{15}}\\
...,
\end{array}
\end{equation}
we have found it advantageous to find the roots of the polynomials formed from the variable transformation $x = \left( {{\Gamma ^{2n - 1}}} \right)z$ with the common factor ${\Gamma ^{2{n^2} - n}}$ factored out;
namely
\begin{equation}
\begin{array}{*{20}{l}}
{{q_1}\left( z \right) = z - 1}\\
{{q_2}\left( z \right) = {z^2} - \displaystyle\frac{{\left( {1 - {\Gamma ^4}} \right)}}{{\left( {1 - {\Gamma ^2}} \right)}}z + 1}\\
{{q_3}\left( z \right) = {z^3} - \displaystyle\frac{{\left( {1 - {\Gamma ^6}} \right)}}{{\left( {1 - {\Gamma ^2}} \right)}}{z^2} + \displaystyle\frac{{\left( {1 - {\Gamma ^6}} \right)}}{{\left( {1 - {\Gamma ^2}} \right)}}z - 1}\\
{{q_4}\left( z \right) = {z^4} - \displaystyle\frac{{\left( {1 - {\Gamma ^8}} \right)}}{{\left( {1 - {\Gamma ^2}} \right)}}{z^3} + \displaystyle\frac{{\left( {1 - {\Gamma ^6}} \right)\left( {1 - {\Gamma ^8}} \right)}}{{\left( {1 - {\Gamma ^2}} \right)\left( {1 - {\Gamma ^4}} \right)}}{z^2} - \displaystyle\frac{{\left( {1 - {\Gamma ^8}} \right)}}{{\left( {1 - {\Gamma ^2}} \right)}}z + 1}\\
{...}
\end{array}
\end{equation}
These polynomials are palindromic/anti-palindromic (depending on order), and possess a closed form
\begin{equation}
{{q_n}\left( z \right) = \sum\limits_{k = 0}^n {q_n^{\left( k \right)}{z^k}} },
\end{equation}
with $q_n^{\left( n \right)} = 1$ and 
\begin{equation}
q_n^{\left( {n - k} \right)} = {{\left( { - 1} \right)}^k}\,\displaystyle\frac{{\prod\limits_{j = 1}^k {\left( {1 - {\Gamma ^{2n + 2 - 2j}}} \right)} }}{{\prod\limits_{j = 1}^k {\left( {1 - {\Gamma ^{2j}}} \right)} }} = {{\left( { - 1} \right)}^k}\left[ \begin{array}{*{20}{c}}
n\\
k
\end{array} \right]_{\Gamma ^2},
\end{equation}
where the Gaussian binomial (or q-binomial) coefficient satisfies the Pascal triangle relation
\begin{equation}
{\left[ {\begin{array}{*{20}{c}}
n\\
k
\end{array}} \right]_y} = {\left[ {\begin{array}{*{20}{c}}
{n - 1}\\
k
\end{array}} \right]_y} + {y^{n - k}}{\left[ {\begin{array}{*{20}{c}}
{n - 1}\\
{k - 1}
\end{array}} \right]_y}.
\end{equation}

In general, the roots of these polynomials all lie on the unit circle in the complex plane (for repulsive interactions), and appear as conjugate pairs. The odd order polynomials have a trivial root of $z=1$, allowing the synthetic factorization into a palindromic polynomial of less degree. In fact, analytic expressions can be obtained for the roots through $g=16$ (polynomial degree $n=9$) \cite{LINDSTROM2015} (see Appendix B).

\begin{figure}[htbp]
\begin{center}
\includegraphics[scale=0.5]{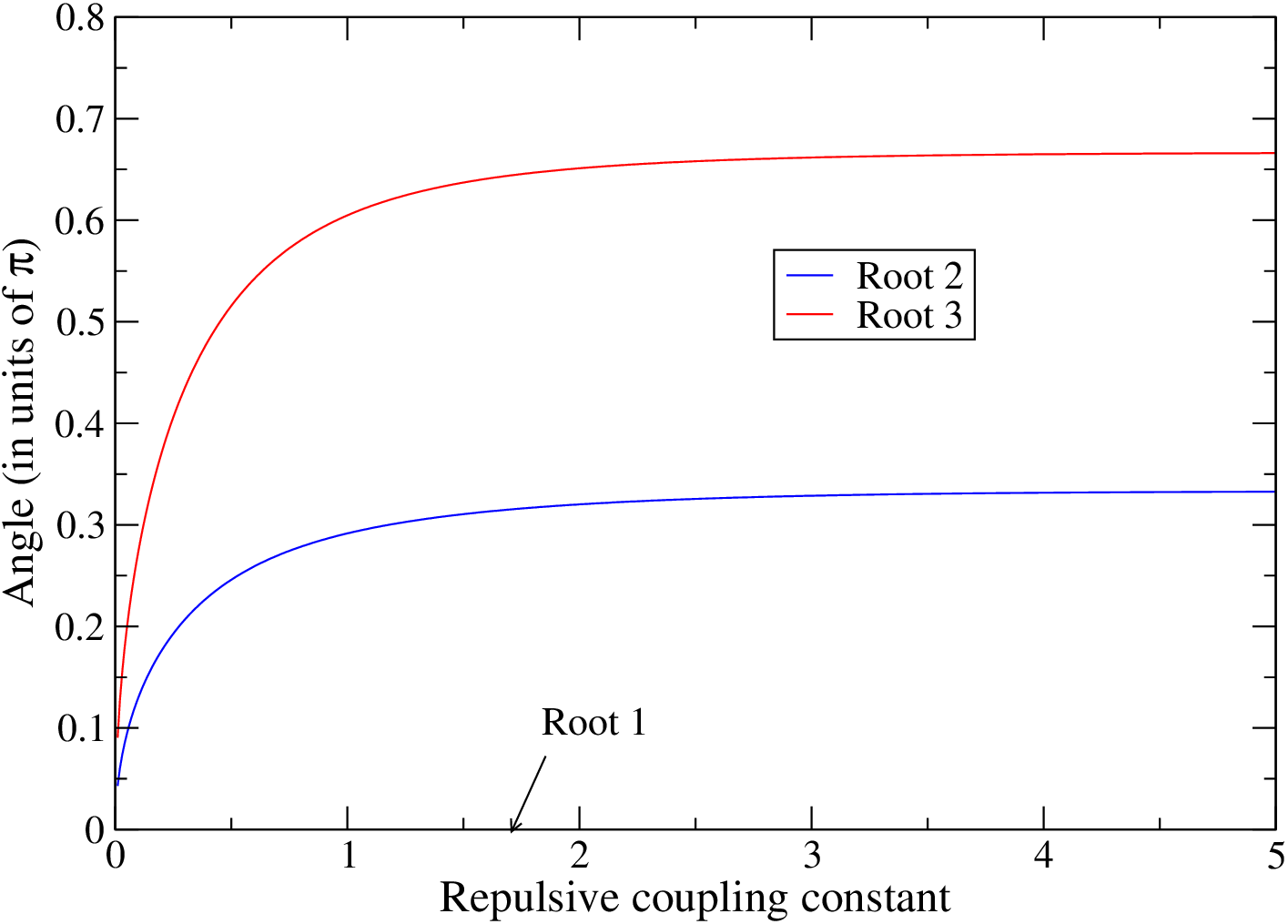}
\caption{Location of roots of the $n=5$ order anti-palindromic potential for the case $g=8$, given by the angle on the unit circle in the complex plane, as a function of the repulsive coupling coefficient. One root is purely real unity (angle zero) while the fourth and fifth root are complex conjugates of roots two and three. The roots for Gaussian quadrature are given upon multiplication by ${{\Gamma ^{2n - 1}}}$.}
\label{fig:root5_loc}
\end{center}
\end{figure}

\begin{figure}[htbp]
\begin{center}
\includegraphics[scale=0.5]{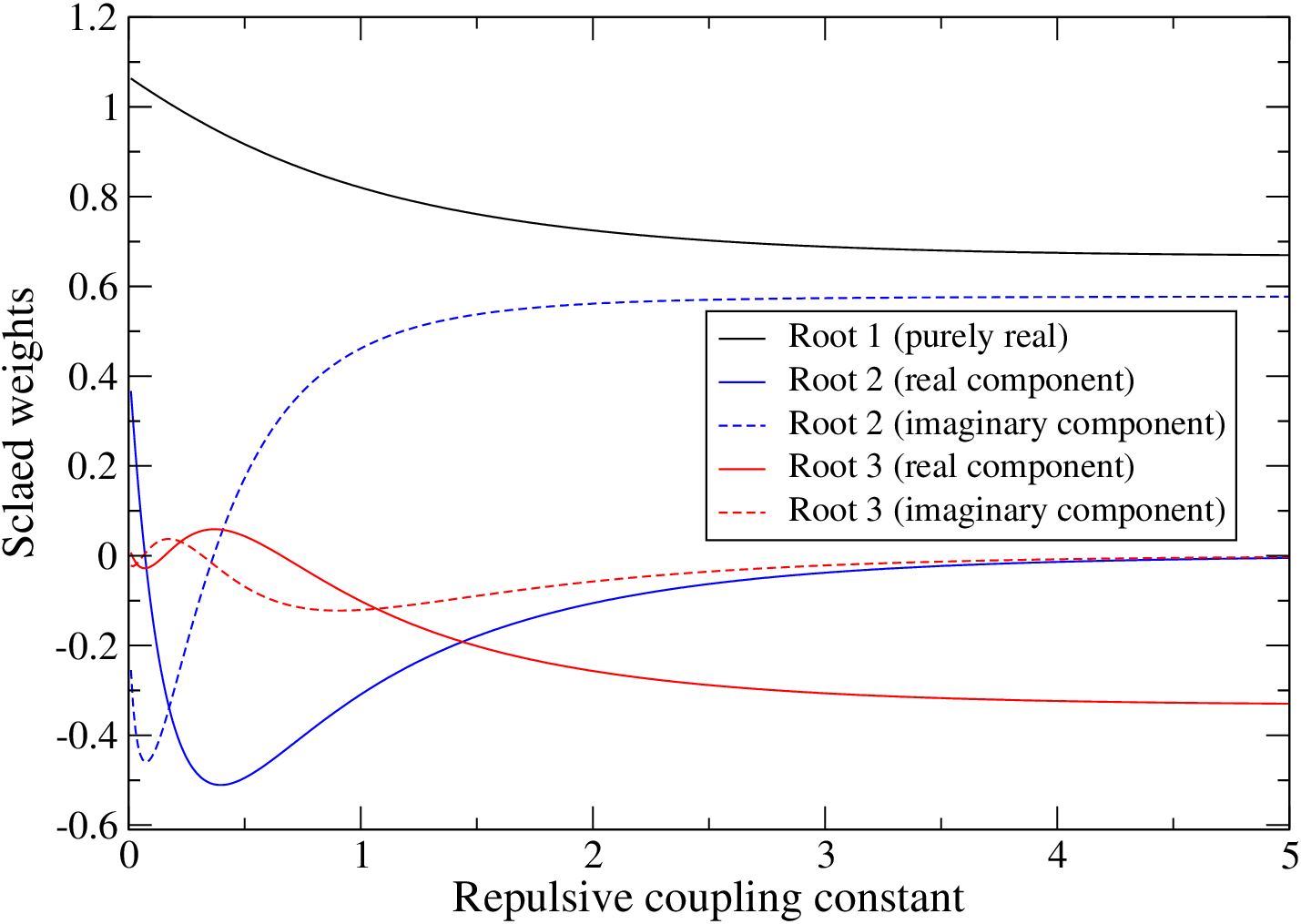}
\caption{Gaussian quadrature weights for the case $g=8$ as a function of the repulsive coupling coefficient. Weights for the fourth and fifth roots are complex conjugates to those of roots two and three. Plotted values were obtained by multiplying with an exponentially decreasing scale factor of ${2{\Gamma^{20}}}$.}
\label{fig:root5_wt}
\end{center}
\end{figure}

Upon obtaining the roots, the weights are obtained in principle from Eq. (\ref{26}). However, for large $g$ and/or large interaction strength (exponentially decreasing $\Gamma$), it is easily seen that the complex weights can grow exponentially large. This is illustrated in
Fig.s \ref{fig:root5_loc} and \ref{fig:root5_wt} for the case $g=8$. Figure \ref{fig:root5_loc} displays the angle of the roots of the $n=5$ order anti-palindromic polynomial on the unit circle in the complex plane potential for the case $g=8$, as a function of the repulsive coupling coefficient. Figure \ref{fig:root5_wt} shows the quadrature weights, still as a function of the repulsive coupling coefficient, weights for the fourth and fifth roots being complex conjugates to those of roots two and three.

As the (complex valued) weights are required to sum to (real valued) unity (from the zero occupation constraint) this signals an un-avoidable loss of numerical precision when using fixed precision computing platforms. In order to be fully general we have surmounted this obstacle by employing arbitrary numerical precision software \cite{BAILEY2022} to implement our algorithms.
 
To find polynomial roots we employed Aberth's algorithm \cite{BINI1996} in double precision to initialize a refinement to arbitrary multiple precision using Laguerre's method \cite{NUMERICALRECIPES}. Generally less than ten iterations of Laguerre's refinement suffices for 400 digit precision. The weights are then found by a straight forward multiple precision evaluation of Eq. (\ref{26}).

\section{Implementation} 

In order to illustrate simply the implementation of discrete Hubbard-Stratonovich transformations let us consider a trivial system consisting of a single super-shell made of two interacting sub-shells of degeneracy $g=2$. The partition function for this system is
\begin{equation}
{U_Q} = \mathop {\sum\limits_{{p_{1 = 0}}}^2 {\;\;\sum\limits_{{p_2} = 0}^2 {} } }\limits_{{p_1} + {p_2} = Q} \quad \left( {\begin{array}{*{20}{c}}
2\\
{{p_1}}
\end{array}} \right)\left( {\begin{array}{*{20}{c}}
2\\
{{p_2}}
\end{array}} \right){e^{ - \beta \left\{ {\left( {{\varepsilon _1} - \mu } \right){p_1} + \left( {{\varepsilon _2} - \mu } \right){p_2} + \frac{1}{2}{\Delta _{11}}{p_1}\left( {{p_1} - 1} \right) + {\Delta _{12}}{p_1}{p_2} + \frac{1}{2}{\Delta _{22}}{p_2}\left( {{p_2} - 1} \right)} \right\}}}.
\end{equation}
Using 
\begin{equation}
{e^{ - \beta {\Delta _{12}}{p_1}{p_2}}} = {e^{ - \frac{1}{4}\beta {\Delta _{12}}\left\{ {{{\left( {{p_1} + {p_2}} \right)}^2} - {{\left( {{p_1} - {p_2}} \right)}^2}} \right\}}} = {e^{ - \frac{1}{4}\beta {\Delta _{12}}{{\left( {{p_1} + {p_2}} \right)}^2}}}{e^{ + \frac{1}{4}\beta {\Delta _{12}}{{\left( {{p_1} - {p_2}} \right)}^2}}},
\end{equation}
let us introduce four interaction parameters
\begin{equation}
{\Gamma _{11}} = {e^{ - \frac{1}{2}\beta \,{\Delta _{11}}}},\quad {\Gamma _{22}} = {e^{ - \frac{1}{2}\beta \,{\Delta _{22}}}},\quad {\Gamma _{12}} = {e^{ - \frac{1}{4}\beta \,{\Delta _{12}}}}\quad \mathrm{and}\quad{\mathord{\buildrel{\lower3pt\hbox{$\scriptscriptstyle\frown$}} 
\over \Gamma } _{12}} = {e^{ + \frac{1}{4}\beta \,{\Delta _{12}}}}
\end{equation}
and four ``auxiliary field'' summations: three for $g=2$ (corresponding to $p_1^2$, $p_2^2$, 
and ${\left( {{p_1} - {p_2}} \right)^2}$), whose $\Gamma$ dependent weights and roots we designate by
${\omega _i}\left( \Gamma \right),{\phi _i}\left( \Gamma \right)$, and one for $g=4$ (corresponding to ${\left( {{p_1} + {p_2}} \right)^2}$), whose weights and roots we designate by
${\mathord{\buildrel{\lower3pt\hbox{$\scriptscriptstyle\frown$}} 
\over \omega } _i}\left( \Gamma \right),{\mathord{\buildrel{\lower3pt\hbox{$\scriptscriptstyle\frown$}} 
\over \phi } _i}\left( \Gamma \right)$.
The partition interacting partition function then can be evaluated exactly as
\begin{equation}
{U_Q} = \sum\limits_{a = 1}^2 {{\omega _a}\left( {{\Gamma _{11}}} \right)} \sum\limits_{b = 1}^2 {{\omega _b}\left( {{\Gamma _{22}}} \right)} \sum\limits_{c = 1}^2 {{{\mathord{\buildrel{\lower3pt\hbox{$\scriptscriptstyle\frown$}} 
\over \omega } }_c}\left( {{\Gamma _{12}}} \right)} \sum\limits_{d = 1}^4 {{\omega _d}\left( {{{\mathord{\buildrel{\lower3pt\hbox{$\scriptscriptstyle\frown$}} 
\over \Gamma } }_{12}}} \right)} \;U_Q^{aux},
\end{equation}
where the auxiliary field partition function is in an independent electron form, and so amenable to exact evaluation
\begin{equation}
U_Q^{aux} = \mathop {\sum\limits_{{p_{1 = 0}}}^2 {\;\;\sum\limits_{{p_2} = 0}^2 {} } }\limits_{{p_1} + {p_2} = Q} \quad \left( {\begin{array}{*{20}{c}}
2\\
{{p_1}}
\end{array}} \right)\left( {\begin{array}{*{20}{c}}
2\\
{{p_2}}
\end{array}} \right)Y_1^{{p_1}}Y_2^{{p_2}},
\end{equation}
with
\begin{equation}
\begin{array}{*{20}{l}}
{{Y_1} = {e^{ - \beta \left( {{\varepsilon _1} - \mu - \frac{1}{2}{\Delta _{11}}} \right)}}{\phi _a}\left( {{\Gamma _{11}}} \right){{\mathord{\buildrel{\lower3pt\hbox{$\scriptscriptstyle\frown$}} 
\over \phi } }_c}\left( {{\Gamma _{12}}} \right){\phi _d}\left( {{{\mathord{\buildrel{\lower3pt\hbox{$\scriptscriptstyle\frown$}} 
\over \Gamma } }_{12}}} \right)}\\
{{Y_2} = {e^{ - \beta \left( {{\varepsilon _2} - \mu - \frac{1}{2}{\Delta _{22}}} \right)}}{\phi _b}\left( {{\Gamma _{22}}} \right){{\mathord{\buildrel{\lower3pt\hbox{$\scriptscriptstyle\frown$}} 
\over \phi } }_c}\left( {{\Gamma _{12}}} \right)\phi _d^{ - 1}\left( {{{\mathord{\buildrel{\lower3pt\hbox{$\scriptscriptstyle\frown$}} 
\over \Gamma } }_{12}}} \right)}.
\end{array}
\end{equation}

\begin{table}[!ht]
\begin{center}
    \begin{tabular}{ccc}\hline\hline
        Sub-shell & Degeneracy & Energy (eV)\\\hline
        3s$_{1/2}$ & 2 & -220.04736\\
        3p$_{1/2}$ & 2 & -180.60865\\
        3p$_{3/2}$ & 4 & -176.99515\\
        3d$_{3/2}$ & 4 & -109.55812\\
        3d$_{5/2}$ & 6 & -108.94330\\
        4s$_{1/2}$ & 2 & -17.234577\\
        4p$_{1/2}$ & 2 & -5.0438986\\
        4p$_{3/2}$ & 4 & -4.4357069\\\hline\hline
    \end{tabular}
\end{center}
\caption{Energy and degeneracy of sub-shells for a copper plasma at $T$=100 eV and $\rho$=8.96 g/cm$^3$ (chemical potential $\mu$=-200.1260 eV) obtained from an average-atom calculation (see Ref. \cite{WILSON1993}).}\label{tab1}
\end{table}

\begin{table}[!ht]
\begin{center}
    \begin{tabular}{ccccccccc}\hline\hline
           & 3s$_{1/2}$ & 3p$_{1/2}$ & 3p$_{3/2}$ & 3d$_{3/2}$ & 3d$_{5/2}$ & 4s$_{1/2}$ & 4p$_{1/2}$ & 4p$_{3/2}$ \\\hline
3s$_{1/2}$ & 21.820 & 22.145 & 22.006 & 22.698 & 22.661 & 8.2991 & 6.8463 & 6.6249 \\        
3p$_{1/2}$ &        & 22.498 & 22.354 & 23.034 & 22.996 & 8.3642 & 6.8989 & 6.6746 \\         
3p$_{3/2}$ &        &        & 22.212 & 22.877 & 22.839 & 8.3420 & 6.8804 & 6.6568 \\
3d$_{3/2}$ &        &        &        & 23.943 & 23.899 & 8.4267 & 6.9482 & 6.7240 \\
3d$_{5/2}$ &        &        &        &        & 23.856 & 8.4204 & 6.9429 & 6.7190 \\
4s$_{1/2}$ &        &        &        &        &        & 4.5828 & 3.8507 & 3.7397 \\
4p$_{1/2}$ &        &        &        &        &        &        & 3.2409 & 3.1483 \\
4p$_{3/2}$ &        &        &        &        &        &        &        & 3.0584 \\\hline\hline
    \end{tabular}
\end{center}
\caption{Interaction matrix $V_{ij}$ (in eV) for sub-shells 3s$_{1/2}$ to 4p$_{3/2}$ in the same conditions as table \ref{tab1} (see Ref. \cite{WILSON1993}).}\label{tab2}
\end{table}

\begin{figure}[!ht]
    \begin{center}
        \includegraphics[scale=0.5]{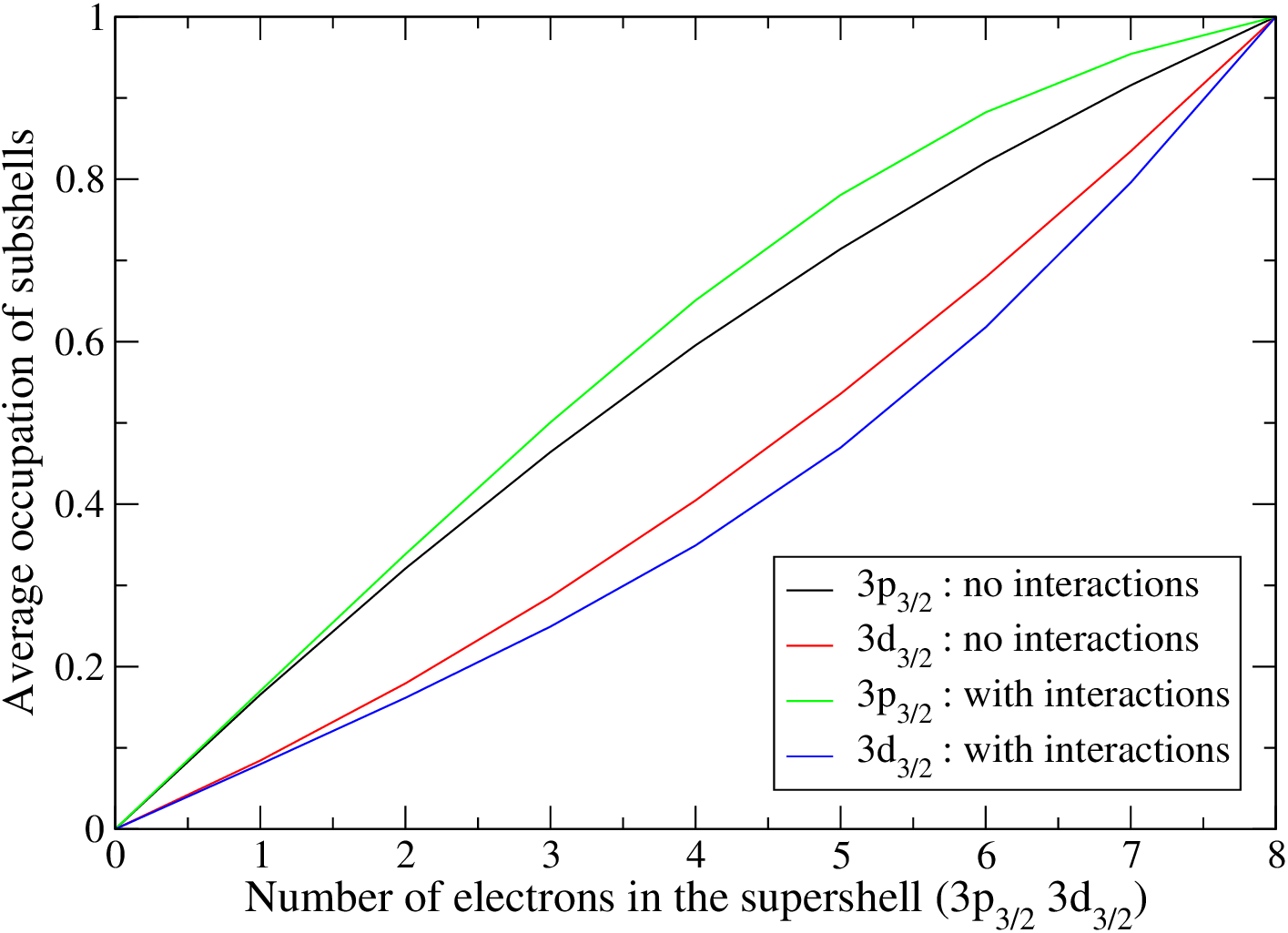}
    \end{center}
    \caption{Average occupation (population divided by degeneracy) of sub-shells 3p$_{3/2}$ and 3d$_{3/2}$ as a function of the number of electrons $Q$ in the case of super-configuration (3p$_{3/2}$ 3d$_{3/2}$)$^Q$.}\label{fig1}
\end{figure}

\begin{figure}[!ht]
    \begin{center}
        \includegraphics[scale=0.5]{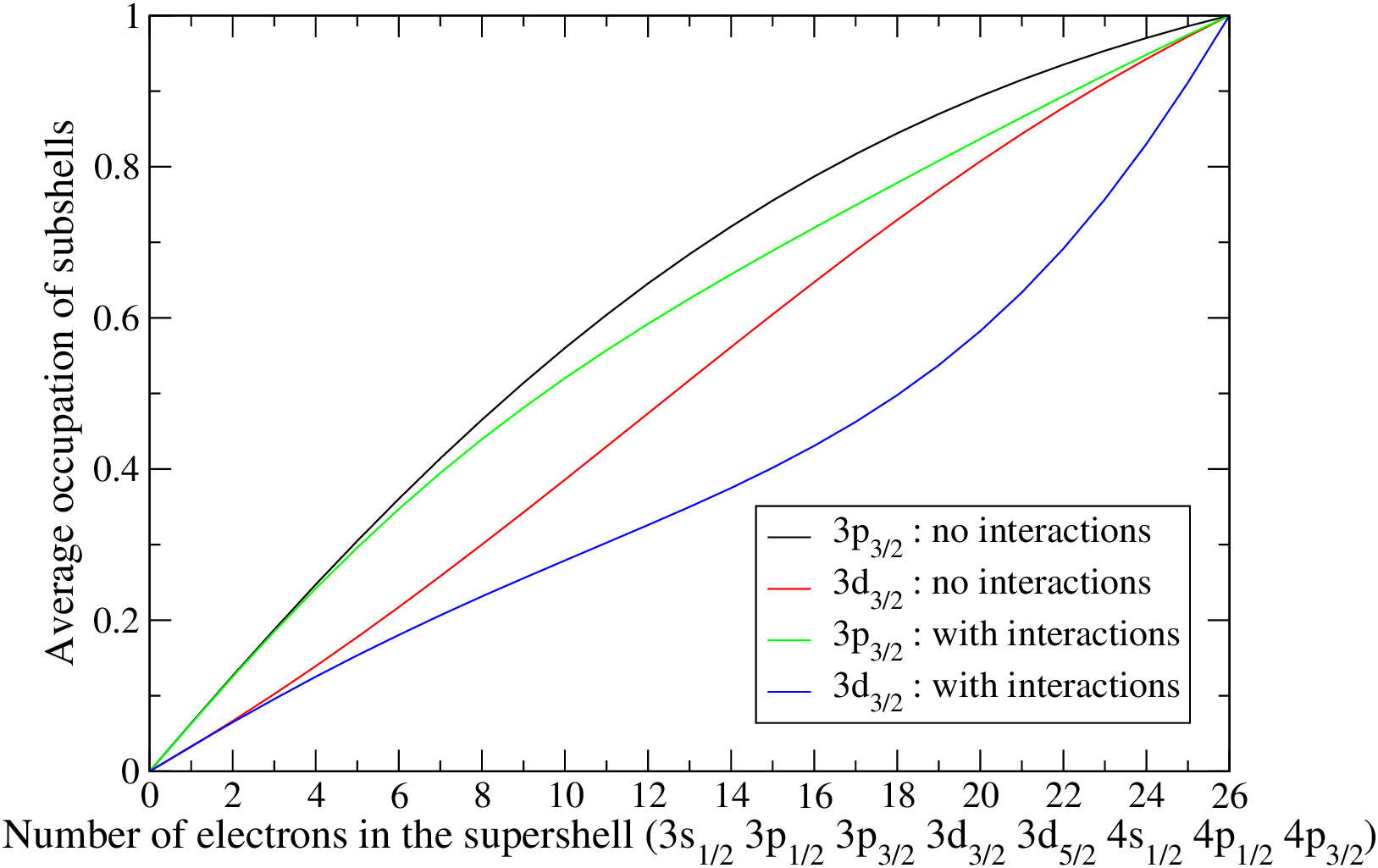}
    \end{center}
    \caption{Average occupation (population divided by degeneracy) of sub-shells 3p$_{3/2}$ and 3d$_{3/2}$ as a function of the number of electrons $Q$ in the case of super-configuration (3s$_{1/2}$ 3p$_{1/2}$ 3p$_{3/2}$ 3d$_{3/2}$ 3d$_{5/2}$ 4s$_{1/2}$ 4p$_{1/2}$ 4p$_{3/2}$)$^Q$.}\label{fig2}
\end{figure}

Since the approach presented here is exact, we recover perfectly the values obtained by a direct brute-force summation. It is instructive to compute the average populations (consisting in setting $A(\vec{p})=p_i$ in Eq. (\ref{avera}), $i$ denoting a given sub-shell) of two sub-shells 3p$_{3/2}$ and 3d$_{3/2}$ when they belong to the large super-shell (3s$_{1/2}$ 3p$_{1/2}$ 3p$_{3/2}$ 3d$_{3/2}$ 3d$_{5/2}$ 4s$_{1/2}$ 4p$_{1/2}$ 4p$_{3/2}$), and when they are just together as (3p$_{3/2}$ 3d$_{3/2}$). The energies of all the considered sub-shells obtained from an average-atom calculation (see Ref. \cite{WILSON1993}) are provided in table \ref{tab1} and the interaction terms in table \ref{tab2}. In the second case (super-shell (3p$_{3/2}$ 3d$_{3/2}$)), the interaction matrix is simply given by table \ref{tab3}. We can see in Fig. \ref{fig1} that the interaction terms can be responsible of around 10 \% of the average population of 3d$_{3/2}$. In the case of super-shell (3s$_{1/2}$ 3p$_{1/2}$ 3p$_{3/2}$ 3d$_{3/2}$ 3d$_{5/2}$ 4s$_{1/2}$ 4p$_{1/2}$ 4p$_{3/2}$), this difference can amount to 35 \% (see Ref. \ref{fig2}). In addition, the impact of the parent super-shell on the effect of interaction terms is very significant, as illustrated by the comparison of Figs. \ref{fig1} and \ref{fig2}. 

\begin{table}[!ht]
\begin{center}
    \begin{tabular}{ccc}\hline\hline
           & 3p$_{3/2}$ & 3d$_{3/2}$ \\\hline        
3p$_{3/2}$ & 22.212 & 22.877 \\
3d$_{3/2}$ &        & 23.943 \\\hline\hline
    \end{tabular}
\end{center}
\caption{Interaction matrix $V_{ij}$ (in eV) for sub-shells 3p$_{3/2}$ and 3d$_{3/2}$ in the case of a copper plasma at $T$=100 eV and $\rho$=8.96 g/cm$^3$ (chemical potential $\mu$=-200.1260 eV) obtained from an average-atom calculation (see Ref. \cite{WILSON1993}).}\label{tab3}
\end{table}

\begin{table}[!ht]
\begin{center}
\begin{tabular}{ccccc}\hline\hline
Configuration & $E^{(0)}$ (eV) & $E^{(1)}$ (eV) & $E^{(2)}$ (eV) & $E_{\mathrm{tot}}$ (eV) \\\hline        
$C_1$=3s$_{1/2}^2$ 3p$_{1/2}^2$ 3p$_{3/2}^4$ 3d$_{3/2}^4$ 3d$_{5/2}^6$ 4s$_{1/2}^2$ 4p$_{1/2}^2$ 4p$_{3/2}^4$ \;\;\;\; & -2663.48 & 705.262 & 3947.97 & 1989.75\\
$C_2$=3s$_{1/2}^2$ 3p$_{1/2}^2$ 3p$_{3/2}^4$ 3d$_{3/2}^4$ 3d$_{5/2}^6$ 4s$_{1/2}^1$ 4p$_{1/2}^1$ 4p$_{3/2}^2$ \;\;\;\; & -2632.33 & 682.146 & 3378.83 & 1428.64\\
$C_3$=3s$_{1/2}^1$ 3p$_{1/2}^1$ 3p$_{3/2}^2$ 3d$_{3/2}^2$ 3d$_{5/2}^3$ 4s$_{1/2}^1$ 4p$_{1/2}^1$ 4p$_{3/2}^2$ \;\;\;\; & -1331.74 & 120.781 & 980.992 & -223.969\\\hline\hline
\end{tabular}
\end{center}
\caption{Different contributions to the energy of three configurations.}\label{tab4}
\end{table}
Finally, the different contributions to the energy $E_{\mathrm{tot}}$ of three configurations are given in table \ref{tab4}. The first contribution represents the linear part
\begin{equation}
E^{(0)}=\sum_{i=1}^Np_i\epsilon_i,    
\end{equation}
the second one the self-interaction quadratic term
\begin{equation}
E^{(1)}=\frac{1}{2}\sum_{i=1}^Np_i(p_i-1)V_{ii},      
\end{equation}
and the third one the other quadratic terms (interactions between electrons belonging to different sub-shells):
\begin{equation}
E^{(2)}=\sum_{i=1, i>j}^Np_ip_jV_{ij}=\frac{1}{2}\sum_{i,j=1}^Np_ip_jV_{ij}.      
\end{equation}
As concerns the three configurations, $C_1$ corresponds to the case where all the sub-shells are full, $C_2$ has the first five sub-shells full and the last three ones half-filled, and in $C_3$ all the sub-shells are half-filled (maximum of the complexity). The linear terms is of course always negative (bound states), and the interaction terms positive. Among the latter, the interaction between non-equivalent electrons overcomes the self-interaction terms (equivalent electrons) by a factor of 5 for $C_1$ and $C_2$, and a factor 8 in the case of $C_3$. The total energies of $C_1$ and $C_2$ are positive, but the energy of $C_3$ is negative. This is due to the fact that the interaction terms vary in a quadratic manner with the sub-shell populations and thus, even if the magnitude of the interaction matrix elements $V_{ij}$ is smaller than the one-electron energies of the first 5 sub-shells, the interaction terms increase faster with the numbers of electrons.

Two generalizations emerge from this trivial example. First, the number of auxiliary field summations scales as the number of sub-shells in the super-shell squared. Second, if ${{p_1}}$ and ${{p_2}}$ belonged to separate super-shells, although the auxiliary partition function could still be factored into independent partition functions for each super-shell, they would still be coupled through the auxiliary field variable summations.

This feature might be considered as too computationally burdensome. One may then wish to consider partitioning the interacting electron energy expression into terms within super-shells, with interaction terms between super-shells treated as a first order correction via a Feynman-Jensen inequality. This hybrid approach would be in-exact, but offer an improvement over current treatments. Likewise, one may wish to treat only the (dominant) diagonal sub-shell to sub-shell interactions via auxiliary fields. It should be remembered, in such approaches, that the reference ideal independent electron energies used for the evaluation of the partition functions would have a different functional form reflecting the new definition of the first order correction factor. The idea of treating only the diagonal terms with the method (and the others by Jensen-Feynman) is interesting because it enables one to use the recursive computation of partition functions, since the diagonal part of a given sub-shell enters the exponential terms as a correction to the one-electron energy. 

The formalism is exact, and enables one to use the machinery of ``independent-particle'' partition functions used in all STA codes, were interaction terms are often averaged for each super-configuration. Making brute-force computation of partition functions (and moments) with interactions terms would require to rewrite entirely STA codes, and would anyway probably be numerically intractable for large super-shells. In addition, some quantities (weights, roots, ...) can be pre-tabulated.

\section{Conclusions}

A new discrete Hubbard-Stratonovich transformation has been introduced for the exact evaluation of interacting electron partition functions, albeit at the cost of additional discrete auxiliary field summations, and the necessity of using arbitrary multiple precision arithmetic to guarantee generality to any sub-shell degeneracy or interaction strength. Due to the multitude of parameter combinations that arise in any given opacity calculation, a general conclusion of the impact on accuracy of such an improved treatment must be relegated to future publications, most probably on a case by case basis.

\vspace{1cm}

\noindent {\bf Acknowledgments}

\vspace{5mm}

This work performed under the auspices of the U.S. Department of Energy by Lawrence Livermore National Laboratory under Contract DE-AC52-07NA27344.

\appendix

\section{Solutions for roots and weights}

Defining $\Gamma \equiv {e^{{\alpha \mathord{\left/ {\vphantom {\alpha 2}} \right. \kern-\nulldelimiterspace} 2}}}$, which is less than
(/greater than) unity for repulsive(/attractive) interactions, we have for $g=2$:
\begin{equation}
{\phi _{1,2}} = \frac{1}{2}\left( {{\Gamma ^3} + {\Gamma ^5} \pm {\Gamma ^3}\Lambda } \right)\quad ;\quad \Lambda = \sqrt { - 3 + 2{\Gamma ^2} + {\Gamma ^4}}
\end{equation}
and
\begin{equation}
{\omega _{1,2}} = \frac{1}{2}\left( {1 \mp \displaystyle\frac{{\left\{ {2 + {\Gamma ^2}} \right\}\Lambda }}{{3{\Gamma ^2} + {\Gamma ^4}}}} \right),
\end{equation}
for $g=4$:
\begin{equation}
{\phi _{1,2}} = \frac{1}{2}\left( {{\Gamma ^7} + {\Gamma ^9} \pm {\Gamma ^5}\Lambda } \right) \quad ; \quad {\phi _3} = {\Gamma ^5}
\end{equation}
as well as
\begin{equation}
{\omega _{1,2}} = \displaystyle\frac{4}{{{\Gamma ^6}\left( {2 + {\Gamma ^2}} \right)\left( {2 + {\Phi _ \pm }} \right)\left( { - 2 + \left( {{\Gamma ^2} + {\Gamma ^4}} \right){\Phi _ \pm }} \right)}} \quad ; \quad {\omega _3} = \frac{{1 + {\Gamma ^2}}}{{{\Gamma ^6}\left( {2 + {\Gamma ^2}} \right)}}\quad ;
\end{equation}
\begin{equation}
\Lambda = \sqrt { - 4 + {\Gamma ^4} + 2{\Gamma ^6} + {\Gamma ^8}} \quad\mathrm{and} \quad {\Phi _ \pm } = {\Gamma ^2} + {\Gamma ^4} \pm \Lambda,
\end{equation}
while for $g=6$ we have expressions for the roots in terms of nested radicals:
\begin{equation}
{\phi _{1,2}} = \displaystyle\frac{\Gamma^7}{4}\left( {{q_a} - {q_b} \mp 2{\Omega _ - }} \right)\quad ; \quad  {\phi _{3,4}} = \displaystyle\frac{\Gamma^7}{4}\left( {{q_a} + {q_b} \mp 2{\Omega _ + }} \right)\quad ;
\end{equation}

\begin{equation}
{q_a} = 1 + {\Gamma ^2} + {\Gamma ^4} + {\Gamma ^6}\quad ; \quad {q_b} = \sqrt {5 - 2{\Gamma ^2} - 5{\Gamma ^4} - {\Gamma ^8} + 2{\Gamma ^{10}} + {\Gamma ^{12}}}
\end{equation}
and
\begin{equation}
{\Omega _ \pm } = \sqrt { - 2 - {q_c} + \displaystyle\frac{q_a^2}{2} \pm \displaystyle\frac{{{q_a}}}{{2{q_b}}}\left( {8 + q_a^2 - 4{q_c}} \right)} \quad \mathrm{and} \quad {q_c} = 1 + {\Gamma ^2} + 2{\Gamma ^4} + {\Gamma ^6} + {\Gamma ^8},
\end{equation}
while the weights (in principle also analytic) are more efficiently evaluated by means discussed below.

We note that for attractive interactions ($\Gamma \ge 1$) the roots and weights are all real, while in the repulsive case they are in general complex valued. These non-linear solution sets are obtainable with computer algebraic programs such as {\sc Mathematica} \cite{MATHEMATICA}, but higher orbital degeneracies could not be obtained in closed form.

\section{Roots of (anti)-palindromic polynomials}

For $g=2$ (polynomial order $n=2$) and $g=4$ (order $n=3$) the roots of the polynomial (the latter factoring out the trivial root $z=1$) 
\begin{equation}
{z^2} - {\alpha _{2,1}}z + 1\quad \mathrm{with}\quad {\alpha _{n,m}} \equiv \frac{{1 - {\Gamma ^{2n}}}}{{1 - {\Gamma ^{2m}}}}
\end{equation}
and
\begin{equation}
\left\{ {z - 1} \right\}\left\{ {{z^2} - \left( {{\alpha _{3,1}} - 1} \right)z + 1} \right\}
\end{equation}
are easily obtained by the quadratic formula. Similarly roots of the polynomials for $g=6$ and $g=8$ (the latter after factoring), namely
\begin{equation}
{z^4} - {\alpha _{4,1}}{z^3} + {\alpha _{4,1}}{\alpha _{3,2}}{z^2} - {\alpha _{4,1}}z + 1
\end{equation}
and
\begin{equation}
{z^4} - \left( {{\alpha _{4,1}} - 1} \right){z^3} + \left( {{\alpha _{4,1}}{\alpha _{3,2}} - {\alpha _{4,1}} + 1} \right){z^2} - \left( {{\alpha _{4,1}} - 1} \right)z + 1
\end{equation}
are solved by quartic formula. 

For $g=10$ and $g=12$ (after factoring) we need to consider the general ${6^{th}}$ degree palindromic polynomial
\begin{equation}
P(z) = {z^6} + {a_1}{z^5} + {a_2}{z^4} + {a_3}{z^3} + {a_2}{z^2} + {a_1}z + 1.
\end{equation}
This polynomial has six roots, which we denote ${r_1},{r_2},{r_3},\frac{1}{{{r_1}}},\frac{1}{{{r_2}}},\frac{1}{{{r_3}}}$ and if we define
\begin{equation}
{\chi _1} = {r_1} + \frac{1}{{{r_1}}},\quad {\chi _2} = {r_2} + \frac{1}{{{r_2}}}\quad \mathrm{and}\quad{\chi _1} = {r_3} + \frac{1}{{{r_3}}},
\end{equation}
we can rewrite our polynomial as
\begin{equation}
P(z) = {z^6} - {s_1}{z^5} + \left( {{s_2} + 3} \right){z^4} - \left( {{s_3} + 2{s_1}} \right){z^3} + \left( {{s_2} + 3} \right){z^2} - {s_1}z + 1,
\end{equation}
where we introduced the elementary symmetric polynomials in the three ``variables'' ${\chi _1},{\chi _2},{\chi _3}$:
\begin{equation}
\begin{array}{l}
{s_1} = {\chi _1} + {\chi _2} + {\chi _3} = - {a_1},\\
{s_2} = {\chi _1}{\chi _2} + {\chi _1}{\chi _3} + {\chi _2}{\chi _3} = {a_2} - 3,\\
{s_3} = {\chi _1}{\chi _2}{\chi _3} = - {a_3} - 2{s_1} = 2{a_1} - {a_3}.
\end{array}
\end{equation}
Now ${\chi _1},{\chi _2},{\chi _3}$ are the three (analytically obtainable) solutions of the cubic equation
\begin{equation}
{z^3} - {s_1}{z^2} + {s_2}z - {s_3} = 0
\end{equation}
and once we find ${\chi _1},{\chi _2},{\chi _3}$ we easily find ${r_1},{r_2},{r_3}$ by solving
\begin{equation}
r_i^2 + {\chi _i}{r_i} + 1 = 0.
\end{equation}
Analytic solutions for $g=14$ and $g=16$ proceed in an analogous fashion by introducing elementary symmetric polynomials in four variables, which require the solution of a quartic equation.

\end{document}